
\font\titlefont = cmr10 scaled\magstep 4
\font\myfont    = cmr10 scaled\magstep 2
\font\sectionfont = cmr10
\font\littlefont = cmr5
\font\eightrm = cmr8 

\def\ss{\scriptstyle} 
\def\sss{\scriptscriptstyle} 

\newcount\tcflag
\tcflag = 0  

\ifnum\tcflag = 0 \magnification = 1200 \fi  

\global\baselineskip = 1.2\baselineskip
\global\parskip = 4pt plus 0.3pt
\global\abovedisplayskip = 18pt plus3pt minus9pt
\global\belowdisplayskip = 18pt plus3pt minus9pt
\global\abovedisplayshortskip = 6pt plus3pt
\global\belowdisplayshortskip = 6pt plus3pt

\def\barsoff{\overfullrule=0pt}


\def\endignore{}
\def\ignore #1\endignore{}

\newcount\dflag
\dflag = 0


\def\monthname{\ifcase\month
\or January \or February \or March \or April \or May \or June%
\or July \or August \or September \or October \or November %
\or December
\fi}

\newcount\dummy
\newcount\minute  
\newcount\hour
\newcount\localtime
\newcount\localday
\localtime = \time
\localday = \day

\def\advanceclock#1#2{ 
\dummy = #1
\multiply\dummy by 60
\advance\dummy by #2
\advance\localtime by \dummy
\ifnum\localtime > 1440 
\advance\localtime by -1440
\advance\localday by 1
\fi}

\def\settime{{\dummy = \localtime%
\divide\dummy by 60%
\hour = \dummy
\minute = \localtime%
\multiply\dummy by 60%
\advance\minute by -\dummy
\ifnum\minute < 10 
\xdef\spacer{0} 
\else \xdef\spacer{} 
\fi %
\ifnum\hour < 12 
\xdef\ampm{a.m.} 
\else 
\xdef\ampm{p.m.} 
\advance\hour by -12 %
\fi %
\ifnum\hour = 0 \hour = 12 \fi
\xdef\timestring{\number\hour : \spacer \number\minute%
\thinspace \ampm}}}



\def\endtitle{}
\def\title#1\endtitle{\vskip.5in\titlefont
\global\baselineskip = 2\baselineskip
#1\vskip.4in
\baselineskip = 0.5\baselineskip\rm}
 
\def\endauthors{}
\def\authors#1\endauthors{#1}

\def\endabstract{}
\def\abstract#1\endabstract{\vskip .3in%
\centerline{\sectionfont\bf Abstract}%
\vskip .1in
\noindent#1}

\def\nopageonenumber{\footline={\ifnum\pageno<2\hfil\else
\hss\tenrm\folio\hss\fi}}  

\newcount\nsection
\newcount\nsubsection

\def\section#1{\global\advance\nsection by 1
\nsubsection=0
\bigskip\noindent\centerline{\sectionfont \bf \number\nsection.\ #1}
\bigskip\rm\nobreak}

\def\subsection#1{\global\advance\nsubsection by 1
\bigskip\noindent\sectionfont \sl \number\nsection.\number\nsubsection)\
#1\bigskip\rm\nobreak}


\def\appendix#1#2{\bigskip\noindent%
\centerline{\sectionfont \bf Appendix #1.\ #2}
\bigskip\rm\nobreak}


\newcount\nref
\global\nref = 1

\def\therefs{} 


\def\ref#1#2{\xdef #1{[\number\nref]}
\ifnum\nref = 1\global\xdef\therefs{\item{[\number\nref]} #2\ }
\else
\global\xdef\oldrefs{\therefs}
\global\xdef\therefs{\oldrefs\vskip.1in\item{[\number\nref]} #2\ }%
\fi%
\global\advance\nref by 1
}

\def\listrefs{\vfill\eject\section{References}\therefs}


\newcount\nfoot
\global\nfoot = 1

\def\foot#1#2{\xdef #1{(\number\nfoot)}
\footnote{${}^{\number\nfoot}$}{\eightrm #2}
\global\advance\nfoot by 1
}


\newcount\nfig
\global\nfig = 1
\def\thefigs{} 

\def\figure#1#2{\xdef #1{(\number\nfig)}
\ifnum\nfig = 1\global\xdef\thefigs{\item{(\number\nfig)} #2\ }
\else
\global\xdef\oldfigs{\thefigs}
\global\xdef\thefigs{\oldfigs\vskip.1in\item{(\number\nfig)} #2\ }%
\fi%
\global\advance\nfig by 1 } 

\def\fig#1{\xdef #1{(\number\nfig)}
\global\advance\nfig by 1 } 


\newcount\ntab
\global\ntab = 1

\def\table#1{\xdef #1{\number\ntab}
\global\advance\ntab by 1 } 


\newcount\cflag
\newcount\nequation
\global\nequation = 1
\def\eqlabel{(1)}

\def\nexteqno{\ifnum\cflag = 0
\global\advance\nequation by 1
\fi
\global\cflag = 0
\xdef\eqlabel{(\number\nequation)}}

\def\lasteqno{\global\advance\nequation by -1
\xdef\eqlabel{(\number\nequation)}}

\def\label#1{\xdef #1{(\number\nequation)}
\ifnum\dflag = 1
{\escapechar = -1
\xdef\draftname{\littlefont\string#1}}
\fi}

\def\clabel#1#2{\xdef\eqlabel{(\number\nequation #2)}
\global\cflag = 1
\xdef #1{\eqlabel}
\ifnum\dflag = 1
{\escapechar = -1
\xdef\draftname{\string#1}}
\fi}

\def\cclabel#1#2{\xdef\eqlabel{#2)}
\global\cflag = 1
\xdef #1{\eqlabel}
\ifnum\dflag = 1
{\escapechar = -1
\xdef\draftname{\string#1}}
\fi}


\def\eeq{}

\def\eqnn #1\eeq{$$ #1 $$}

\def\eq #1\eeq{
\ifnum\dflag = 0
{\xdef\draftname{\ }}
\fi 
$$ #1
\eqno{\eqlabel \rlap{\ \draftname}} $$
\nexteqno}



\def\eol{& \eqlabel \rlap{\ \draftname} \crcr
\nexteqno
\xdef\draftname{\ }}



\def\eeolnn{\xdef\draftname{\ }}

\def\eqa #1\eeq{
\ifnum\dflag = 0
{\xdef\draftname{\ }}
\fi 
$$ \eqalignno{ #1 } $$
\global\cflag = 0}


\def\ie{{\it i.e.\/}}
\def\eg{{\it e.g.\/}}
\def\etc{{\it etc.\/}}

\def\via{{\it via\/}}



\global\nulldelimiterspace = 0pt



\def\frac#1#2{{{#1} \over {#2}}\,}  
\def\hf{{1\over 2}}
\def\nth#1{{1\over #1}}


\def\Asl{\hbox{/\kern-.7500em\it A}} 
\def\Dsl{\hbox{/\kern-.6700em\it D}} 
\def\dsl{\hbox{/\kern-.5300em$\partial$}}
\def\pxpsl{\hbox{/\kern-.5600em$p$}}
\def\sslsh{\hbox{/\kern-.5300em$s$}}
\def\epssl{\hbox{/\kern-.5100em$\epsilon$}}
\def\delsl{\hbox{/\kern-.6300em$\nabla$}}
\def\lxpsl{\hbox{/\kern-.4300em$l$}}
\def\elxpsl{\hbox{/\kern-.4500em$\ell$}}
\def\kxpsl{\hbox{/\kern-.5100em$k$}}
\def\qxpsl{\hbox{/\kern-.5000em$q$}}
\def\sla#1{\raise.15ex\hbox{$/$}\kern-.57em #1}



\def\roughly#1{\mathrel{\raise.3ex\hbox{$#1$\kern-.75em\lower1ex\hbox{$\sim$}}}}
\def\lsim{\roughly<}



\def\bfj{{\bf j}}


\def\Bfa{{\bf A}}


\def\Scl{{\cal L}}


\def\ssb{{\sss B}}

\def\sse{{\sss E}}

\def\ssm{{\sss M}}

\def\ssq{{\sss Q}}

\def\ssS{{\sss S}}
\def\sst{{\sss T}}


\def\diag#1{{\rm diag}\left( #1 \right)}



\def\Avg#1{\left\langle #1 \right\rangle}






\nopageonenumber
\baselineskip = 18pt
\barsoff


\font\myfont    = cmr10 scaled\magstep 2
\def\BigTc{$\hbox{\titlefont T}_{\hbox{\myfont c}}$}

\def\Tc{{$T_c$}}
\def\five{{$SO(5)$}}
\def\four{{$SO(4)$}}
\def\threetwo{{$SO(3) \times SO(2)$}}
\def\vfive{{E_5}}
\def\vexp{{E_{\rm sb}}}
\def\dt{{\partial_t}}
\def\nq{{n_\ssq}}
\def\ns{{n_\ssS}}
\def\min{{\rm min}}
\def\max{{\rm max}}
\def\AF{{\sss AF}}
\def\SC{{\sss SC}}
\def\MX{{\sss MX}}


\rightline{November, 1996.}
\line{hep-th/9611070 \hfil McGill-96/45, OSLO-TP **-96}

\title
\centerline{On the SO(5) Effective Field} 
\centerline{Theory of High \BigTc\ Superconductors}
\endtitle

\vskip 0.1in
\authors
\centerline{C.P. Burgess${}^a$ and C.A. L\"utken${}^b$}  
\vskip .2in
\centerline{\it ${}^a$ Physics Department, McGill University}
\centerline{\it 3600 University St., Montr\'eal, Qu\'ebec, Canada, H3A 2T8.}
\vskip .1in
\centerline{\it ${}^b$ Physics Department, University of Oslo}
\centerline{\it P.O. Box 1048, Blindern, N-0316 Norway.}
\endauthors

\footnote{}{\eightrm * Research supported by N.S.E.R.C. of Canada, 
F.C.A.R. du Qu\'ebec, and the Norwegian Research Council.}


\abstract
\vbox{\baselineskip 15pt 
We construct the low-energy effective theory for the $SO(5)$ 
model of high-$T_c$ superconductivity, recently proposed by 
S.C. Zhang (cond-mat/9610140). This permits us to develop a 
systematic expansion for low-energy observables in powers of 
the small symmetry-breaking interactions. The approximate 
$SO(5)$ symmetry predicts relations amongst these observables, 
which are model-independent consequences of Zhang's proposed 
symmetry-breaking pattern.}   
\endabstract


\vfill\eject

\section{Introduction}

\ref\zhang{S.-C. Zhang, {\it SO(5) Quantum Nonlinear $\sigma$ Model
Theory of the High $T_c$ Superconductivity}, Stanford
preprint, cond-mat/9610140.}

In a recent remarkable paper \zhang\ Zhang has suggested an elegant framework
for understanding the relationship between the superconducting (SC) and
antiferromagnetic (AF) properties of the high-\Tc\ superconductors. His proposal
comes in two parts. First, he argues that these systems enjoy an 
approximate \five\ symmetry which contains as a subgroup the \threetwo\ symmetry
of spin rotations and electromagnetic gauge transformations. The \five\ symmetry
is only approximate, in the sense that it is explicitly broken to \threetwo\ by
small interaction terms in the Hamiltonian, whose scale we denote by $\vexp$. It
is also explicitly broken by the doping of electrons or holes away from half
filling. Second, the \five\ symmetry is argued to be spontaneously broken to
\four\ by the dynamics which binds the electrons into spin-singlet pairs. The
energy scale, $\vfive$, which characterizes the order parameter for this
breaking is imagined to be comparable to the Fermi energy, and is taken much
larger than $\vexp$. For energies much smaller than $\vfive$, the
symmetry-breaking dynamics is purely concerned with how the \five-breaking order
parameter aligns relative to the direction of explicit symmetry breaking. In
ref.~\zhang\ Zhang argues convincingly how such a framework can synthesize a
great many features of these systems, and proposes a model low-energy effective 
theory which qualitatively describes their properties, and does so quantitatively in
the vicinity of the critical point between the antiferromagnetic (AF) and 
superconducting (SC) regimes.

\ref\pseudos{S. Weinberg, Phys. Rev. Lett. {\bf 29}, 1698 (1972)}

\ref\chiral{S. Weinberg, Phys. Rev. Lett. {\bf 18}, 188 (1967);
Phys. Rev. {\bf 166}, 1568 (1968); Physica {\bf 96A}, 327 (1979);
C.G. Callan, S. Coleman, J. Wess and B. Zumino, Phys. Rev. {\bf 177}, 2247 (1969);
J. Gasser and H. Leutwyler, Ann. Phys. {\bf 158}, 142 (1984);
Nucl. Phys. {\bf B250}, 465 (1985)}

Here we generalize Zhang's model to incorporate the {\it most general} possible
interactions which are consistent with the assumed symmetry-breaking pattern.
Being the most general possible such lagrangian, it {\it must} then incorporate
the low-energy limit of any particular microscopic model which realizes these
symmetries. Because the low-energy behaviour involves the interactions of
Goldstone (and pseudo-Goldstone\foot\pGBs{A pseudo-Goldstone boson is the
Goldstone boson for a symmetry which is only approximate.} \pseudos) bosons,
they are guaranteed to interact only weakly at low energies. Consequently, a
mean-field treatment of the effective theory is justified to describe these
systems for energies well below $\vfive$, and away from any critical points.
This description closely parallels that of chiral perturbation theory \chiral,
which has been very successfully used to describe the low-energy interactions of
pions and nucleons within the framework of Quantum Chromodynamics.  

\ref\toappear{C.P. Burgess and C.A. L\"utken, in preparation.}

Eq.~(5) expresses our main results, which consist of a number of relations
amongst the low-energy observables of these systems which are generic
consequences of, and therefore strong tests of, the approximate \five\ symmetry. For
the sake of brevity we here simply state our results, with a more thorough discussion
to appear elsewhere.

\ref\mynotes{For an introduction to Goldstone bosons in high-energy and
condensed matter physics, see: C.P. Burgess, {\it An Introduction to Effective
Lagrangians and their Applications}, lecture notes for the Swiss Troisi\`eme
Cycle, Lausanne, June 1995.}


We start with the lagrangian density for the four would-be Goldstone bosons,
$\theta^\alpha$, $\alpha = 1,\dots,4$, corresponding to the spontaneous
breakdown $SO(5) \to SO(4)$ \mynotes. Writing this as a derivative expansion,
and, for simplicity, ignoring the system's spatial anisotropy
gives:\foot\dimensions{We choose units for which $\ss \hbar = c = k_\ssb
= \vfive = 1$.}    
\label\invform
\eq
\Scl_{\rm inv}(\theta) = - V_0 + {f^2_t \over 2} \; \dt {n}^\sst \dt {n} - 
{f^2_s \over 2} \; \nabla {n}^\sst \cdot \nabla {n} + \hbox{higher derivatives}.
\eeq
Here $V_0$, $f_t$ and $f_s$ are real constants which are of order unity (in
units where $\vfive = 1$). The four variables, $\theta^\alpha$, parameterize the
ground-state field configurations, which fill out the space $SO(5)/SO(4)$ or,
equivalently, the 4-sphere. We use in \invform\ coordinates consisting of a
five-dimensional vector --- the `superspin' --- ${n}(\theta)$, having unit
length, ${n}^\sst {n} = 1$. $n$ transforms simply with respect to \five: ${n}
\to g \, {n}$, for $g$ a real, orthogonal five-by-five matrix. 

In order to explicitly break \five\ down to \threetwo\ we introduce the
symmetry-breaking matrix: $M = \epsilon \, \diag{m_q,m_s,m_s,m_s,m_q}$. Here
$\epsilon = \vexp/\vfive \sim$ few \%, is the small dimensionless parameter
which describes the strength of the explicit \five\ breaking. Following Zhang,
we choose the spin $SO(3)$ to rotate $n_2, n_3$ and $n_4$ into one another,
while the electromagnetic $SO(2)$ rotates $n_1$ into $n_5$. The explicit
breaking of \five\ symmetry by doping is achieved within the microscopic theory
by coupling a chemical potential, $\mu$, to the electron's electric charge,
$Q$. In units of $\vfive$, we expect $\mu$ to lie in the range $|\mu| \lsim 0.1
- 0.2$ for the doping of high-\Tc\ systems. Since electric charge is a generator
of $SO(5)$, we take: $Q_{ij} = q (\delta_{1i} \, \delta_{5j} - \delta_{1j} \,
\delta_{5i})$. $q = \pm 2$ is the electric charge (in units of $e$) of the
superconducting order parameter. 

The lagrangian for the system is then the most general local function of
$n(\theta)$, $M$ and $\mu Q$ which satisfies $\Scl(g{n},g M g^\sst, g \mu Q
g^\sst) = \Scl({n}, M, \mu Q)$. The leading terms in a derivative expansion
therefore are: 
\label\generalform
\eqa
\Scl_{\rm sb} &=  - V + f_t^2 \, \Bigl[ A \dt \nq^\sst
\, \dt \nq + B \, \dt \ns^\sst \, \dt \ns + C \, (\nq^\sst \,
\dt \nq)^2 \Bigr] \eol 
& \qquad - f_s^2 \,\Bigl[ D \nabla \nq^\sst \cdot \nabla \nq - E 
\, \nabla \ns^\sst \cdot \nabla \ns - F \, (\nq^\sst \, \nabla \nq)^2 \Bigr] 
+ \hbox{higher derivatives}, \eeolnn  
\eeq
where $\nq$ and $\ns$ respectively are the components of $n$ within the
electromagnetic and spin subspaces. The functions $V$, $A$, $B$, $C$, $D$, $E$
and $F$ are all functions of $n$, $M$ and $\mu Q$. For instance the scalar
potential, $V$, has the general form:    
\label\generalfnform 
\eq
V[n,M,\mu Q] = \sum_{k=1}^\infty 2^{-k} (v_{k\ssq} \, \nq^\sst \nq
+ v_{k\ssS} \, \ns^\sst \ns )^k .
\eeq
Analogous expansions may also be written for the functions $A$ through $F$ (with
coefficients denoted $a_{k\ssq}, a_{k\ssS}$ through $f_{k\ssq}, f_{k\ssS}$).
Coupling to electromagnetic fields, $A_\mu$, is achieved by replacing ordinary
derivatives of $\nq$ with covariant ones: $\partial_\mu \nq \to D_\mu \nq =
\partial_\mu \nq - q e A_\mu Q \, \nq$.

It is useful to express the conserved spin and electromagnetic currents, which
are obtained from \invform\ and \generalform\ using Noether's theorem, for the
effective theory, since it is to these that external probes often couple. The
terms involving the fewest derivatives are:
\label\invcurrents
\eq
\eqalign{
\rho_{\rm em} = -  f_t^2 \, \Bigl(1 + 2A \Bigr) \nq^\sst \dt\nq , \qquad 
&\bfj_{\rm em} = f_s^2 \, \Bigl( 1 + 2D \Bigr) \nq^\sst Q \nabla \nq , \cr 
\vec{\rho}_{\rm spin} = f_t^2 \, \Bigl( 1 + 2B \Bigr) \vec{\ns}
\times \dt \vec{\ns} , \qquad
&\vec{\bfj}_{\rm spin} = - f_s^2 \, \Bigl( 1 + 2E \Bigr) \vec{\ns} \times \nabla
\vec{\ns}  . \cr}  
\eeq

Notice that eqs.~\invform, \generalform\ and \invcurrents\ also capture the form
for the free energy and currents at nonzero temperature if all coefficients are
understood as functions of $T$. This is because these expressions are the most
general possible consistent with the derivative expansion and the assumed
symmetry-breaking pattern. 

The predictive power of eq.~\generalform\ emerges once we perturb in the small
symmetry-breaking parameters $\epsilon$ and $\mu$. This requires knowing the
dependence of the coefficients in eq.~\generalfnform\ (and its analogs for the
functions $A$ to $F$) on these small parameters. It is an easy exercise to see
that the following properties hold: $v_{k\ssS}, a_{k\ssS}, b_{k\ssS}, c_{k\ssS},
d_{k\ssS}, e_{k\ssS}$ and $f_{k\ssS}$ are independent of $\mu$. These
coefficients are at most $O(\epsilon)$ in size, as are all of the others when
restricted to half-filling: $\mu=0$. The antisymmetry of the matrix $Q$ further
implies that $v_{k\ssq}, a_{k\ssq}, b_{k\ssq}, d_{k\ssq}$ and $e_{k\ssq}$ must
be {\it even} under the interchange $\mu \to -\mu$. The same is {\it not} true
for the coefficients $c_{k\ssq}$ and $f_{k\ssq}$. Because most low-energy
observables do not appreciably depend on these last coefficients, they also
enjoy a symmetry between positive and negative doping. 

Perturbing in $\epsilon$ and $\mu$ is powerful because all interactions amongst
the $\theta^\alpha$ in eqs.~\invform\ and \generalform\ vanish in the limit
$\epsilon = \mu = 0$. Mean field theory is therefore justified so long as three
criteria hold: $(i)$ $\mu$ and $\epsilon$ are small; $(ii)$ the derivative
expansion is justified (long wavelengths compared to $1/\vfive$; and $(iii)$ 
we are not in the vicinity of critical points, for which non-Goldstone modes
become massless, with the associated uncontrollable fluctuations in the
infrared. 


We now summarize some of the predictions which follow to leading order in
$\epsilon$ and $\mu$. For details concerning their derivation, and some results
beyond leading order, see \toappear. To leading order it suffices to take
$A=B=C=D=E=F=0$ and keep only the \five-invariant interactions having two
derivatives. The symmetry-breaking terms then enter predictions only through the
potential, $V$. It suffices to keep only the terms having $k=1,2$ in the
expansion \generalfnform, and to expand the coefficients $v_{1\ssq}$ and
$v_{2\ssq}$ to linear order in $\mu^2$: \eg: $v_{1\ssq} = v^0_{1\ssq} +
v^1_{1\ssq} \mu^2$ \etc. In this limit the phase diagram, long-wavelength
dispersion relations, and electric and magnetic stiffness are all determined by
five parameters, which we choose to be $m^2 \equiv v^0_{1\ssq} - v_{1\ssS} =
O(\epsilon)$, $\kappa \equiv - v^1_{1\ssq} = O(1)$, $\xi \equiv (v^1_{2\ssq})^2
= O(1)$, $f_s = O(1)$ and $f_t = O(1)$. 

Since these five parameters relate more than five low energy observables for
the SC and AF phases, we obtain testable relationships amongst these
observables, which we now summarize. 


Writing the magnitudes of $\ns$ and $\nq$ as $|\ns| = \sin\theta$ and $|\nq| =
\cos\theta$, and minimizing the potential with respect to $\theta$ gives two
generic phases: $\theta_\min = 0, \pi$ (SC); and $\theta_\min = {\pi \over 2},
{3\pi \over 2}$ (AF). The curvature of the potential at these extrema are   
$M^2_\AF \equiv \left( {d^2 V / d\theta^2} \right)_{\theta = {\pi\over 2}}
\approx  m^2 - \kappa \mu^2$, and $M^2_\SC \equiv \left( {d^2V / d\theta^2}
\right)_{\theta = 0} \approx - m^2 + \kappa \mu^2 - \xi \mu^4$. Notice the sum
rule: $M^2_\AF + M^2_\SC = - (v^1_{2\ssq} \mu^2)^2 \le 0$, which guarantees that
either or both of these curvatures must be negative (for all $\mu$). A third,
mixed, phase (MX) --- Zhang's `spin-bag' phase --- exists when {\it both}
$M^2_\AF$ and $M^2_\SC$ are negative. In this case the minima are $\cos
2\theta_\MX = (M^2_\SC - M^2_\AF)/(v_{2\ssq} - v_{2\ssS})^2$, and the curvature
at this minimum is $M^2_\MX = -2 \, M^2_\SC M^2_\AF / (M^2_\SC + M^2_\AF)$.

The critical doping which defines the boundary between these phases is found
by solving the conditions $M^2_\AF(\mu^2_\AF) = 0$ and $M^2_\SC(\mu^2_{\SC\pm})
= 0$, giving $\mu^2_\AF = \mu^2_{\SC-} + O(\epsilon^2) = ({m^2 /\kappa}) +
O(\epsilon^2)$, and $\mu^2_{\SC+} = ({\kappa / \xi}) + O(\epsilon)$. We see
that the AF phase is predicted to extend only for dopings which are
$O(\epsilon)$, \ie\ a few \%. The mixed phase, whose existence requires
$\mu^2_{\SC-} < \mu^2_\AF$, occurs over a range of dopings which extend to
$O(\epsilon^2)$ past the AF phase boundary, if it exists at all. Then one enters
the SC phase, which persists over a wide $O(1)$ range of dopings. These results
break down for $\mu \sim \mu_{\SC+}$, since there the expansion in powers of
$\mu^2$ and $1/\vfive$ fail.

For temperatures satisfying $T_0(\mu) < T < \vfive$ \five\ remains broken, but
thermal transitions are possible between the SC and AF phases. 
An estimate for $T_0(\mu)$ 
is obtained by finding the height of the potential barrier in $V$ as
a function of $\mu$. For $\mu$ chosen so that it is the AF phase which minimizes
$V$, we find: $T_0(AF) \propto V_{\rm barrier} = \nth{4} ( M^2_\AF - M^2_\SC )
\approx \hf \, (m^2 - \kappa \mu^2)$. On the other hand, choosing $\mu$ so that
$V$ favours the SC phase gives: $T_0(SC) \propto V_{\rm barrier} = \nth{4} (
M^2_\SC - M^2_\AF)$.


As discussed by Zhang, in each phase the four modes, $\theta^\alpha$, group
themselves into {\it bona fide} gapless Goldstone modes (having dispersion
relation $\omega^2(k) = c^2_0 \, k^2$ at low energies), and pseudo-Goldstone
states (for which $\omega^2(k) = c^2_p \, k^2 + \varepsilon^2$). The
lagrangian given by \invform\ and \generalform\ predicts the observables $c_0$,
$c_p$ and $\varepsilon$ for each phase. For brevity we record here only the
lowest-order predictions for the AF and SC phases. 

In the AF phase two modes are gapless (magnons), with the other two forming
an electrically charged pseudo-Goldstone doublet with respect to the unbroken
$SO(2)$. We find $(c_0)_\AF = (c_p)_\AF = c_s \equiv f_s/f_t +
O(\epsilon,\mu^2)$. The gap is given as a function of doping by
$\varepsilon^2_\AF(\mu^2) = (m^2 - \kappa \mu^2)/f^2_t$. 

\ref\oldzhang{E. Demler and S.C. Zhang, Phys. Rev. Lett.{\bf 75}, 4126 (1995)}

\ref\nscattexps{H.A. Mook et al., Phys. Rev. Lett. {\bf 70}, 3490 (1993); 
J. Rossat-Mignod et al., Physica (Amsterdam) {\bf 235C}, 59 (1994);
H.F. Fong et al., Phys. Rev. Lett. {\bf 75}, 316 (1995) }

In the SC phase there is only one gapless mode, which is `eaten' by the photon
\via\ the Anderson-Higgs mechanism, with the remaining three states forming a
spin-triplet pseudo-Goldstone state. It is this state which Zhang argues
persuasively \oldzhang\ has been seen in neutron scattering experiments
\nscattexps\ in the spin-flip channel. The predictions in this case are:
$(c_0)_\SC = (c_p)_\SC = c_s$, and $\varepsilon^2_\SC = (-m^2 + \kappa \mu^2 -
\xi \mu^4)/f_t^2$.  


The final observables for which we present predictions are the electric and
magnetic stiffness in the SC phase. Coupling to an electromagnetic potential,
and eliminating the SC phase's Goldstone mode by an appropriate gauge choice, 
we find the term: $\Scl' = \hf \, a_\sse^2 A_0^2 - \hf \, a_\ssm^2
\Bfa^2$, with $a_\sse = q e f_t$ and $a_\ssm = q e f_s$. $a_\sse$ and $a_\ssm$
represent the medium's electric and magnetic stiffness, normalized so that the
penetration depth of electric and magnetic fields are respectively given by
$a_\sse^{-1}$ and $a_\ssm^{-1}$. 


By eliminating the parameters $m^2$, $\kappa$, $\xi$, $f_t$ and $f_s$
we obtain several parameter-independent relations among the observables
described above. We summarize these here:
\label\predictions
\eq
\eqalign{
\varepsilon^2_\AF(\mu) &= {\varepsilon_\AF^2(0) \over \mu^2_\AF} \, \Bigl[
\mu^2_\AF - \mu^2 \Bigr] ,\cr
\varepsilon^2_\SC(\mu) &= {\varepsilon^2_\SC(\max) \over \mu^4_\max } \, (\mu^2
- \mu^2_{\SC-}) (2\mu^2_\max - \mu^2) ,\cr
{\varepsilon_\AF^2(0) \over \mu^2_\AF} &= 2 \; {\varepsilon^2_\SC(\max) \over
\mu^2_\max } ,\cr
\mu^2_\AF &= \mu^2_{\SC-} + O(\epsilon^2), \cr
(c_0)_\AF &= (c_p)_\AF = (c_p)_\SC = (c_0)_\SC = {a_\ssm \over a_\sse} .\cr} 
\eeq
In these expressions $\varepsilon_\AF^2(0) = m^2/f_t^2$ denotes the AF gap
energy at zero doping, and $\varepsilon^2_\SC(\max) = ({ \kappa^2 / 4 \xi
f_t^2})$ denotes the maximum gap size in the SC phase, which occurs when
$\mu^2_\max = \mu_{\SC+}/2 = \kappa /2 \xi$.

These are robust predictions of the assumed symmetry-breaking pattern {\it
regardless of the nature of the microscopic physics which is believed to be
responsible}, making their experimental verification particularly interesting.
They receive calculable corrections in powers of $\epsilon$, which are most
easily computed using the effective lagrangian presented here \toappear. More
predictions are possible should the MX phase arise, since the properties of the
three Goldstone and one pseudo-Goldstone states in this phase are predictable in
terms of the {\it same} parameters used here. 


We close with a speculation concerning how the \five\ model might account for
the anomalous electrical resistivity of the normal phase of the high-\Tc\
materials, which depend linearly on temperature (when doped with holes and near
optimal doping). Within the \five\ model this resistivity arises within a
`normal' phase below the \five-breaking scale, but at temperatures higher than
the barrier between the AF and SC phases. The characteristic new feature of this
`normal' phase is the existence there of the light pseudo-Goldstone states, some
of which carry electric charge. No such states are present in the normal phase
of a BCS superconductor. In the rest of this section we wish to give a (naive)
argument that if these light states play a significant role as carriers of
electric current, then their scattering might give rise to a resistivity which
is linear in $T$. 

To see why this is so, we estimate the temperature dependence of the resistivity
by computing the temperature-dependence of the two-body rate for scattering of
the appropriate charge carriers from various quasiparticles within the material:
$\rho(T) \sim \Avg{n \sigma v} \sim n \Pi |A|^2 $. Here $n$ describes the
density of `target' quasiparticles from which the charge carriers scatter, while
$A$ describes the amplitude for this scattering and $\Pi$ represents the phase
space available for the final states. The average is over the thermal
distribution of scatterers, and the initial distribution of charge carriers. 
This kind of estimate reproduces the low-temperature $T^5$ dependence when
applied to electron-phonon scattering, and the $T^2$ dependence when applied to
electron-electron scattering. 

The basic observation now comes in two parts. First, notice that the
electrically-charged, pseudo-Goldstone states of Zhang's \five\ model obey a
`relativistic' kinematics since their gap energy is smaller than the
temperatures of interest. Second, unlike low-energy phonons, the
pseudo-Goldstone states need not interact only through derivative couplings,
and so their scattering amplitude, $A$, from other particles need not be
suppressed by powers of $T$ at low temperatures. Combining these observations 
gives the standard estimate for the scattering rate of a `relativistic'
particle, which follows purely on dimensional grounds: $\rho \propto n \Pi
|A|^2  \propto T$. In this, admittedly simplistic, picture a resistivity which
is linear in $T$ is seen as the signature that an appreciable part of the
electromagnetic current is being carried by the light pseudo-Goldstone charge
carriers. This motivates better understanding of the transport properties of
these systems in this temperature range. 

The results of our analysis are encouraging, with the effective theory
naturally accounting for the features of the high-\Tc\ phase diagram. The
predictions of eqs.~\predictions\ make an even more quantitative test of the
\five-invariant picture. Clearly much more remains to be done with this
low-energy lagrangian, most notably including a systematic discussion of their
transport properties, and an analysis of the response of the pseudo-Goldstone
modes to various probes. 

\bigskip
\centerline{\bf Acknowledgments}
\bigskip

We would like to thank the organizers of the Oslo workshop on low-dimensional 
physics (June 1996) for providing the pleasant milieu where this research
started, and Professor Shou-Cheng Zhang for fruitful conversations there
concerning his ideas about high $T_c$ superconductors. This research was
partially funded by the N.S.E.R.C. of Canada, F.C.A.R du Qu\'ebec and the
Norwegian Research Council.

{\baselineskip 15pt
\listrefs
}

\bye